\documentclass[12pt]{article}

\usepackage{newtxtext,newtxmath}

\usepackage{graphicx}
\usepackage{multirow}%

\usepackage{amsthm}
\usepackage{mathrsfs}%
\usepackage[title]{appendix}%
\usepackage{xcolor}%
\usepackage{textcomp}%
\usepackage{manyfoot}%
\usepackage{booktabs}%
\usepackage{algorithm}%
\usepackage{algorithmicx}%
\usepackage{algpseudocode}%
\usepackage{listings}%
\usepackage{gensymb}
\usepackage{ulem}

\usepackage[letterpaper,margin=1in]{geometry}

\renewenvironment{abstract}
	{\quotation}
	{\endquotation}

\date{}

\makeatletter
\renewcommand{\fnum@figure}{\textbf{Figure \thefigure}}
\renewcommand{\fnum@table}{\textbf{Table \thetable}}
\makeatother

\usepackage{scicite}

\usepackage{amsmath,amssymb,amsfonts}
\usepackage{siunitx}
\usepackage{url}

\def\scititle{
	Giant orbital magnetoresistance in the antiferromagnet CoO driven by dynamic orbital angular momentum interaction
}
\title{\bfseries \boldmath \scititle}

\author{
	Christin~Schmitt$^{1}$,
	Sachin~Krishnia$^{1,\ast}$,
    Edgar~Gal\'{i}ndez-Ruales$^{1}$,
    Luca~Micus$^{1}$,\and
    Takashi~Kikkawa$^{2,3}$, 
    Hiroki~Arisawa$^{2,4}$, 
    Marjana~Le\v{z}ai\'c$^{5}$, 
    Duc~Tran$^{1}$, \and
    Timo~Kuschel$^{1,6}$,
    Jairo~Sinova$^{1}$,
    Eiji~Saitoh$^{2,4,7,8}$,
    Gerhard~Jakob$^{1}$, \and
    Olena~Gomonay$^{1}$, 
    Yuriy~Mokrousov$^{1,5}$,
    Mathias~Kl\"aui$^{1,9,\ast}$ \and
	\small$^{1}$Institute of Physics, Johannes Gutenberg-University Mainz, Mainz \& 55128, Germany.\and
	\small$^{2}$Department of Applied Physics, The University of Tokyo, Tokyo \& 113-8656, Japan.\and
    \small$^{3}$Advanced Science Research Center, Japan Atomic Energy Agency, Tokai \& 319-1195, Japan.\and
    \small$^{4}$RIKEN Center for Emergent Matter Science, Wako \& 0351-0198, Japan.\and
    \small$^{5}$Peter Gr\"unberg Institut and Institute for Advanced Simulation, Forschungszentrum J\"ulich and JARA,\and \small J\"ulich \& 52425, Germany. \and
    \and \small$^{6}$Faculty of Physics, Bielefeld University, Bielefeld \& 33615, Germany.\and
    \small$^{7}$WPI-Advanced Institute for Materials Research, Tohoku University, Sendai \& 980-8577, Japan.\and
    \small$^{8}$Institute for AI and Beyond, The University of Tokyo, Tokyo \& 113-8656, Japan.\and
    \small$^{9}$Center for Quantum Spintronics, Norwegian University of Science and Technology, Trondheim \& 7491, Norway.\and
	\small$^\ast$Corresponding authors. Email: skrishni@uni-mainz.de, klaeui@uni-mainz.de\and
}

\begin{document} 

\maketitle

\begin{abstract} \bfseries \boldmath
Recent predictions of orders of magnitude larger orbital current effects compared to spin currents have attracted significant interest. However, the full potential of giant orbital currents remains to be fully harnessed, since so far, the orbital currents need to be converted into spin currents before they can interact with the static magnetization that is dominated by spin angular momentum in conventional magnets. By using a magnet dominated by orbital angular momentum, we demonstrate a more than fifty-fold enhancement in orbital Hall magnetoresistance in CoO/Cu*, compared to conventional CoO/Pt. This is found to be driven by a unique interaction between dynamic orbital angular momentum from surface oxidized Cu* (i.e., the orbital current) and the static orbital angular momentum which constitutes the magnetic moments in the antiferromagnetic insulator CoO. 
A distinctive scattering mechanism for orbital currents at the CoO interface leads to a sign reversal in orbital magnetoresistance in CoO/Cu* compared to CoO/Pt. Our results show how by using orbital angular momentum-dominated materials such as CoO, we can harness the benefits of giant orbital currents that have not been possible using conventional spin-dominated magnets, for orbitronics-based devices, offering unprecedented energy efficiency for operations of antiferromagnets that combine ultimate stability with THz dynamics. 
\end{abstract}

\noindent

\subsection*{Introduction}
In recent decades, the interaction between static magnetization and the flow of spin angular momentum (spin currents), have been employed as the key mechanism to detect and control the magnetic state of ferromagnets. This interaction has enabled major  applications, ranging from giant magnetoresistance sensors~\cite{Baibich1988,Binasch1989} to spin-orbit torque magnetic random-access memory~\cite{miron2011,BHATTI2017}, chiral skyrmion-based racetrack memory~\cite{Fert2013,Everschor2018}, logic devices~\cite{Zhang2015}, and neuromorphic applications~\cite{Grollier2020,gomes2023,Beneke2024}. However, the generation of spin current relies heavily on the spin-orbit coupling (SOC) of the materials, which is a weak relativistic effect that limits the efficiency and constrains materials choices~\cite{Manchon2019}.

More recently,  the dynamic orbital angular momentum (OAM) $-$ the  counterpart of electron spin angular momentum $-$ has emerged as a promising alternative to manipulate magnetization more efficiently~\cite{go2021orbitronics,DongwookPRL2018,Shilei2020,Lee2024}. Two mechanisms of orbital current generation are orbital Hall effect (OHE) in the bulk \cite{Tanaka2008, Kontani2009,DongwookPRL2018} and orbital Rashba Edelstein effect (OREE) at interfaces \cite{Johansson2021,Nikolaev2024}, analogous to those of spin current generation~\cite{Fert2024}. Unlike spin currents, whose generation from charge currents relies on the relatively weak SOC mechanism, orbital currents can be directly generated from charge currents because the crystal momentum of charge carriers (e.g. electrons) can directly couple to OAM. This results in a more robust, orders-of-magnitude larger generation of current-induced OAM (orbital currents), even in light, abundant, inexpensive and environmentally friendly materials~\cite{DongwookPRL2018,salemi2022first,krishnia2023,Choi2023, Kawakami2023, Idrobo2024}. Despite these advantages, the fundamental limitation in utilizing orbital currents for device applications lies in their low interaction efficiency with magnetization that in conventional transition metals is carried by spin angular momentum. While spin currents interact efficiently with magnetization (or spin moments) via strong \textit{s-d} exchange interaction, orbital currents cannot couple in conventional magnetic materials with nearly quenched OAM by any strong exchange interaction and, therefore, one cannot make use of the large orbital currents to effectively interact with conventional spin-based magnets.

To address this issue of inefficient interaction with conventional magnets, the mitigation strategy has been the conversion of orbital currents into spin currents using materials that exhibit high SOC~\cite{Shilei2020,bose2023}. However, the  efficiency in such cases again relies on the comparably weak relativistic SOC mechanism in the conversion layer. Although this approach achieves a modest enhancement in spin torque efficiency, typically a fewfold~\cite{Shilei2020,Krishnia2024,Sala2022,gupta2024}, it remains significantly smaller than the theoretically predicted orders of magnitude enhancement.

Thus, to fully exploit the potential of large orbital currents, one needs to directly couple it with orbital magnetization carried by orbital angular momentum (orbital magnetism), thus bypassing the need for inefficient orbital-to-spin conversion. Recent theoretical studies predict that  orbital-orbital interactions can be a sizable fraction of the \textit{s-d} exchange interaction, indicating that the role of orbital exchange could be significant~\cite{Katsnelson2010}, which bodes well for the exploration of materials with strong orbital magnetism. Antiferromagnetic (AFM) transition metal oxide insulators can exhibit strong orbital magnetization and at the same time, they exhibit unique properties such as THz spin dynamics~\cite{Rongione2023,kampfrath2011coherent}, immunity to external magnetic fields enhancing the stability and low damping, which allows for the transport of angular momentum over long distances \cite{lebrun2018tunable, das}. By making use of the interaction between orbital current from Cu* and unquenched orbital magnetization in CoO, we demonstrate a two orders of magnitude enhancement in the magnetoresistance (MR) in CoO/Cu* bilayers compared to conventional spin-based CoO/Pt. To corroborate that the observed effect originates specifically from the orbital physics and is not a general property of AFM insulators, we measure the  MR in $\alpha$-Fe$_2$O$_3$/Cu* bilayers, for which $\alpha$-Fe$_2$O$_3$ exhibits nearly quenched OAM leading to a low conventional MR response~\cite{rollmann2004first}. Our results show that the full efficiency of OHE-based mechanisms can be harnessed when orbital current interacts with local orbital magnetic momentum as found in insulating AFMs.

\subsection*{Results}

\begin{figure} 
	\centering
	\includegraphics[width=0.9\textwidth]{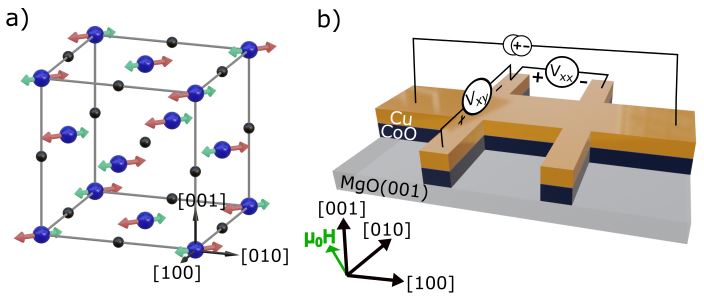} 
	\caption{\textbf{Magnetic structure of CoO and experimental geometry.}
		(\textbf{A}) Illustration of the magnetic and crystallographic structure of CoO with in-plane magneto-crystalline anisotropy, showing the N\'eel vector easy axis along the [$\bar{1}10$] axis. The Co and O atoms are represented by blue and black solid circles, respectively. The red and green arrows denote the spin and orbital angular momenta of each Co atom, respectively, pointing in opposite directions~\cite{satoh2017excitation}. (\textbf{B}) Schematic of the transverse MR measurement scheme with electrical contacts in the CoO/Cu* structure, including the magnetic field sweep direction $[1\bar{1}0]$ shown by a green arrow.}
	\label{fig1} 
\end{figure}

We use CoO, a collinear insulating AFM, for which the OAM of each Co atom is sizeable ($\approx 2.05~\mu_\mathrm{B} /\rm{atom} $)~\cite{norman1990orbital}, unlike in 3\textit{d} transition metal ferromagnets for which orbital moments are nearly fully quenched due to the crystal field ($< 0.1~ \mu_\mathrm{B} /\rm{atom}$)~\cite{CTChenPRL1995,krishnia2024interfacial}. The unquenched OAM 
contributes to both the staggered magnetization and magnetic anisotropies~\cite{grzybowski2023antiferromagnetic}. Thus CoO serves as an ideal system to investigate the interaction between orbital currents and orbital magnetization and to distinguish it from spin magnetization effects. The insulating nature of CoO is particularly suitable as it allows us to rule out effects of current flow in the magnet such as self-torques or anisotropic MR effects~\cite{Wang2019,Krishnia2021}. The collinear magnetic and crystallographic structure of CoO is shown in Figure \ref{fig1}(A).

We deposited epitaxial CoO($\SI{5}{nm}$)/Pt($\SI{2}{nm}$) and CoO($\SI{5}{nm}$)/Cu*($\SI{6}{nm}$) thin films on MgO(001) substrates using reactive magnetron sputtering (see Materials and Methods S1). Here, '*' denotes the natural oxidation of Cu, which serves as an orbital current generator via the OHE and/or OREE \cite{Shilei2020}, while Pt generates spin current via the spin Hall effect mechanism~\cite{miron2011}. As shown in Figure. \ref{fig1}(A), the CoO thin films exhibit four-fold in-plane magneto-crystalline anisotropy with two magnetic easy axes of the N\'eel vector (\textbf{n}) oriented along the $[110]$ and $[\bar{1}10]$ crystallographic axes \cite{baldrati2020efficient, grzybowski2023antiferromagnetic, schmitt2024mechanisms}.  The N\'eel temperature ($T_\mathbf{N}$) of our thin films is approximately $\SI{300}{K}$, slightly higher than that of bulk CoO ($T_\mathbf{N} = \SI{291}{K}$ for bulk CoO~\cite{roth1958magnetic, saito1966x}), which we attribute to the influence of epitaxial strain from the MgO substrate (see Materials and Methods S1 and Supplementary Text S2).

To investigate the interaction between orbital current and static OAM, we fabricated $\SI{6}{\micro m}$ wide Hall bar structures (see Figure \ref{fig1}(B) for the example of CoO/Cu*) and measured orbital Hall magnetoresistance (OMR) (see Materials and Methods S1). When a charge current \textit{I} is injected in the Hall bar, it generates orbital current in Cu* via OHE and/or OREE. An orbital accumulation $\mu_o$ is then built up at the CoO/Cu* interface. Analogous to the spin Hall magnetoresistance (SMR) mechanism~\cite{Nakayama2013}, the OMR refers to the change in electrical resistance of a nonmagnetic material due to the interaction of orbital currents with the magnetization of an adjacent magnetic layer. The change arises from the absorption or reflection of the orbital currents at the interface, depending on the magnetic configuration of the AFM, which in our case is governed by the orientation of the N\'eel vector. 

\begin{figure}[h]
\centering
\includegraphics[width=0.9\textwidth]{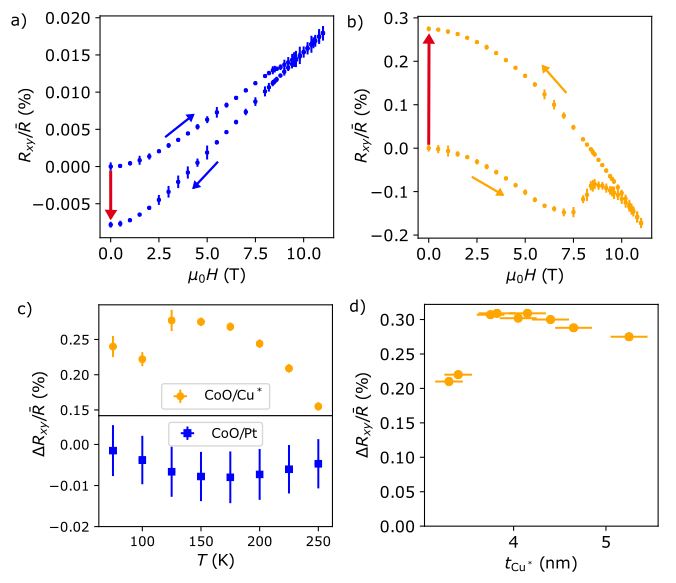}
\caption{\textbf{Spin and orbital Hall magnetoresistance}. Magnetic field-sweep measurements on (\textbf{A}) CoO(\SI{5}{nm})/Pt(\SI{2}{nm}) and (\textbf{B}) CoO(\SI{5}{nm})/Cu*(\SI{6}{nm}) samples, respectively. Both samples exhibit a spin-flop transition between \SI{7.5}{T} - \SI{9}{T}. The change in transverse resistance corresponding to the two perpendicular N\'eel vector orientations is indicated by the red arrow for SMR in \textbf{A} and OMR in \textbf{B}. The measurements are performed at $T = \SI{150}{K}$. (\textbf{C}) The difference in transverse resistance, normalized with the longitudinal resistance for two perpendicular N\'eel vector states at zero field as a function of temperature. The top panel presents the results for CoO/Cu* (in yellow), while the bottom panel shows the results for CoO/Pt (in blue). (\textbf{D}) The difference in transverse resistance for two perpendicular N\'eel vector states at zero field as a function of Cu-layer thickness in CoO/Cu* (t). The measurements were performed at a constant temperature of $T = \SI{150}{K}$.} \label{fig2}
\end{figure}

Although SMR has been studied in Pt adjacent to a range of materials, including 3\textit{d} ferromagnets and insulating magnets \cite{Junyeon2016,Nakayama2013}, the reported pure OMR remains weak for the materials studied~\cite{Ding2022} primarily due to the lack of direct coupling to conventional spin-dominated magnetization in these materials. We first consider a device of CoO/Pt to check the interaction of spin currents with CoO to compare it to the interaction of orbital currents in CoO/Cu*. Initially, we apply an  external magnetic field of $\SI{13}{T}$ along the $[110]$ direction of CoO to align the N\'eel vector along the well-defined $[\bar{1}10]$ magnetic easy axis~\cite{baldrati2020efficient}. We then sweep the magnetic field in $[\bar{1}10]$ direction and probe the orientation  of N\'eel vector by measuring the transverse resistance ($R_{xy}$) at $T (<T_\mathbf{N}) = \SI{150}{K}$. $R_{xy}$ exhibits different behavior when the magnetic field is swept upward and downward in the range of $7.5-\SI{9}{T}$ as shown in Figure \ref{fig2}(A). This change in $R_{xy}$, normalized by the longitudinal resistance, represents the spin-flop transition and rotation of the N\'eel vector by 90$\degree$ i.e. from $[\bar{1}10]$ to $[110]$ in this case. The orientation of the N\'eel vector remains unchanged upon reducing the magnetic field to zero~\cite{baldrati2020efficient}. Therefore, the difference in $R_{xy}$ at zero fields, before and after the magnetic field sweep, effectively probes the two perpendicular N\'eel vector orientations, providing a measure of SMR amplitudes. Furthermore, since the SMR is estimated from resistance measurements taken at zero magnetic fields, additional contributions to the MR  from the metallic layers, such as Hanle magnetoresistance \cite{velez2016hanle}, can be ruled out. The amplitude of SMR ($\Delta R_{xy}/\bar{R}$) in CoO/Pt at $\SI{150}{K}$ is found to be 0.0078$\%$. This amplitude of the SMR is in the order of what is found for all combinations of Pt with magnetic insulators with spin-based magnetization such as NiO, YIG and hematite~\cite{hoogeboom2017negative,
Marmion2014,Nakayama2013}. Based on the theory of SMR in magnetic insulators, the spin Hall angle ($\theta_{\textrm{SH}}$) of 2 nm thick Pt, with a resistivity of $\SI{26}{\micro \Omega cm}$ at $\SI{150}{K}$, is estimated to be $\approx$ 3.5 $\%$ with spin diffusion length $\lambda_s \approx \SI{1.5}{nm}$, which agrees well with  previous studies~\cite{baldrati2020efficient}. Here, we use a literature value for the real part of the spin-mixing conductance to be $G_r$ = $5\times \SI{E14}{\Omega^{-1} m^{-2}}$ \cite{Nakayama2013}.

As the first key result, we now compare this to OMR measurements on the CoO/Cu* sample. Figure \ref{fig2}(B) shows the transverse resistance $R_{xy}$, normalized by the longitudinal resistance, as a function of the magnetic field measured at $T = \SI{150}{K}$ (below $T_\mathbf{N}$). The magnetic field protocol used is the same as that for the measurements in Figure \ref{fig2}(A). An abrupt change in $R_{xy}$ is observed at the spin-flop field, which occurs at approximately 7.5-$\SI{9}{T}$. This value is similar to that observed for CoO/Pt, indicating that the magneto-crystalline anisotropy of CoO remains largely unaffected by the choice of metallic layer on top. The observation of MR in the absence of any significant spin current source indicates the presence of orbital currents originating from Cu*, consistent with previous observations of OMR in Cu* adjacent to metallic ferromagnets~\cite{Ding2022}. Most importantly, we observe a change in the sign of  $R_{xy}$ at the spin-flop field compared to the CoO/Pt structure. This result represents a unique characteristic of orbital current interaction with the unquenched orbital moment in CoO. Such an interaction has not been reported in previous studies on 3d transition metal ferromagnets with quenched OAM~\cite{Ding2022}. The most striking result is the OMR magnitude of 0.28$\%$, which is approximately \textit{36} times larger than the SMR observed in CoO/Pt. 

Although Cu* has been recently identified as an efficient source of orbital current~\cite{Shilei2020,Krishnia2024}, the observed two orders of magnitude higher MR in CoO/Cu* cannot be explained solely by the large orbital currents, as the magnitude of OMR in our previous results for NiFe/Cu* (0.05 $\%$) is found to be significantly smaller than the OMR in our CoO/Cu*~\cite{Ding2022}. Furthermore, torque experiments reveal that the torques on magnetic layers in which the OAM is quenched, are not enhanced by the here observed two orders of magnitude since the orbital currents from Cu* need to be converted into spin currents via the weak SOC of Pt~\cite{Shilei2020,Krishnia2024}. Our experimental results thus suggest that additional factors or mechanisms must contribute to the giant amplitude of the MR in the CoO/Cu* system. Specifically, our findings provide strong evidence that while spin currents from Pt interact with the spin magnetization in CoO via spin exchange coupling, the orbital current from Cu* interacts strongly with the unquenched component of the orbital magnetization in CoO.  This interaction is dominant in CoO/Cu* samples in which no significant source of spin current is present, and is thus responsible for the giant OMR compared to the conventional CoO/Pt. Our experimental findings are consistent with recent theoretical calculations predicting that the orbital-orbital exchange interaction can be comparable in magnitude to spin-spin exchange interaction~\cite{Katsnelson2010}. To validate this interpretation, we compare the results to the MR in $\alpha$-Fe$_2$O$_3$/Cu*, for which the static OAM is quenched and does not contribute significantly to the magnetization of $\alpha$-Fe$_2$O$_3$ (see Supplementary Text S3). The results show that the OMR is nearly absent in the $\alpha$-Fe$_2$O$_3$/Cu* system, and since the insulating nature of the magnet does not allow for orbital to spin conversion, no SMR is expected, thus, supporting our interpretation. 

We next comment on the opposite sign of MR in the two systems. The sign of  spin current from Pt and orbital current from Cu* is theoretically predicted~\cite{salemi2022first} and  experimentally found to be the same, as confirmed by the same sign of SMR and OMR in NiFe/Pt and NiFe/Cu* systems~\cite{Ding2022}. The unquenched OAM of CoO can give rise to an orbital quadrupole moment. Our results indicate that the dynamics of the orbital quadrupole moment in CoO are fundamentally different from the classical dynamics of ferromagnets. A similar anomaly in quadrupole moment dynamics has recently been experimentally demonstrated~\cite{Yoon2023}. 

To understand the underlying mechanisms for the large MR magnitude and its sign, we compare the temperature dependence of the MR in the two systems. Figure \ref{fig2} (C) shows the evolution in MR as a function of temperature in CoO/Cu* (top panel) and CoO/Pt (bottom panel). The sign of MR in CoO/Cu* remains opposite to that of CoO/Pt throughout the measured temperature range, however, the magnitude does vary. In CoO/Cu*, the MR signal increases initially, from 0.15$\%$ at $\SI{275}{K}$ to a peak of $0.28\%$ at $\SI{150}{K}$, followed  by a slight decrease as the temperature is further lowered to $\SI{75}{K}$. For comparison, we also measured the temperature dependence of MR in CoO/Pt, which shows a similar temperature trend, but with an opposite sign and a two-order of magnitude lower amplitude if compared to CoO/Cu*. Remarkably, the amplitude of MR in CoO/Cu* sample compared to the CoO/Pt sample increases further at lower temperatures. At $T = \SI{200}{K}$,  MR in CoO/Cu* is a factor 35 larger than the CoO/Pt. At $T = \SI{100}{K}$ the ratio increases to a factor of \textit{59}. Although a similar temperature dependence has been previously  observed in studies of SMR and torques in YIG/Pt bilayers~\cite{Marmion2014}, as well as in ferromagnetic/Pt bilayers, the increase in the ratio of OMR in CoO/Cu to SMR in CoO/Pt suggests that two different mechanisms are involved. We attribute this temperature dependence primarily to angular momentum relaxation mediated by phonon scattering, which likely plays a significant role~\cite{Marmion2014} for SMR but orbital angular momentum is not prone to these damping mechanisms. Other possible effects, such as spin and/or orbital mixing conductance variations, are considered to be less significant for the observed temperature dependence in these systems, as the interface properties remain constant across the temperature range.

\begin{figure}[h]
\centering
\includegraphics[width=0.45\textwidth]{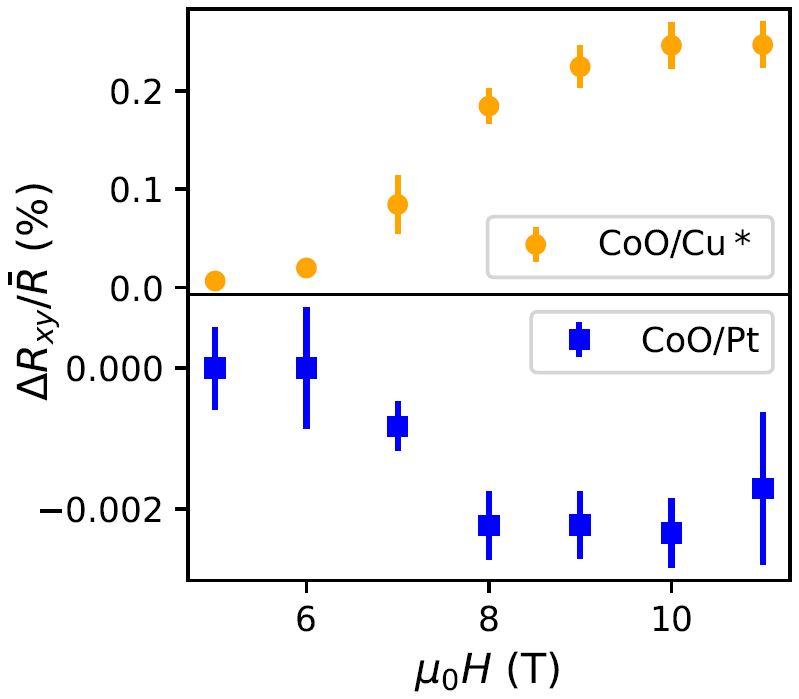}
\caption{\textbf{External field dependence of MR.} The change in transverse MR for CoO(\SI{5}{nm})/Cu*(\SI{6}{nm}) and CoO(\SI{5}{nm})/Pt(\SI{2}{nm}) as a function of applied magnetic field at $T=\SI{200}{K}$.}\label{fig3}
\end{figure}

The observed orbital current could have its origin primarily in either the bulk Cu* via OHE or at the surface via OREE. To determine the origin, we perform Cu* thickness-dependent measurements of OMR in CoO/Cu*(t) samples at $\SI{150}{K}$. To investigate this within the same device, i.e. keeping the CoO/Cu* interface unchanged, we progressively reduce the Cu* layer thickness using a Ar$^+$ ion milling technique, followed by exposure to air for 12 hours after each etching step. The remaining Cu* thickness after each etching step was estimated by comparing the sheet resistance of the milled device to that of separately grown Cu*(t) thin films (see Supplementary Text S4). In Figure 2(D), we show the Cu* layer thickness dependence of the OMR. The results reveal that the OMR nearly remains constant at approximately $\SI{0.32}{\%}$ for Cu* thicknesses ranging from 3.7 to 5.5~nm. When the Cu* thickness is reduced further to 3.2~nm, the OMR decreases slightly, from $\SI{0.32}{\%}$ to $\SI{0.22}{\%}$. For Cu* thicknesses below $\SI{3.2}{nm}$, the sample becomes insulating, making it impossible to measure any OMR. This observed trend suggests that the orbital current responsible for the OMR in our samples predominantly originates at the surface of the oxidized Cu layer, likely facilitated by an OREE. The gigantic enhancement of OMR observed can thus result from the efficient generation of large orbital currents at the oxidized Cu surface by OREE. Furthermore, the nearly constant OMR in the thickness range of 3.7 to $\SI{5.5}{nm}$ implies that this mechanism is robust within this thickness regime, likely due to sufficient Cu volume and thus constant oxygen gradient and Cu* thickness. This supports the orbital current generation and transport and makes it more suitable for applications. 

To determine whether the observed effect originates from the CoO layer and to rule out contributions from other effects, such as the Hanle effect and Lorentzian magnetoresistance in metallic layers, we measure the angular-dependence of $R_{xy}$ for different magnetic field amplitudes. The magnetic field strengths range from values below to above the spin-flop fields of CoO. We extract the MR amplitude as detailed in Supplementary Text S5. Figure \ref{fig3} shows that the amplitude of the MR is nearly zero for magnetic field strengths below the spin-flop transition for both samples. As the magnetic field is increased beyond the spin flop field (8–9 T), the MR signal saturates, confirming that the observed effect is field-independent and ruling out the contribution from other spurious signal origins. The amplitude of the MR  measured is consistent with the values obtained using the zero-field approach (see Figure~\ref{fig2}C) at $T = \SI{200}{K}$, showing the robustness of our measurement approach  and the effect. Additionally, the observed hysteresis in the transverse MR is in line with the presence of significant unquenched OAM in our sample~\cite{grzybowski2023antiferromagnetic}.

\subsection*{Discussion and outlook}

CoO is known for its uncompensated atomic-like orbital magnetism of $d$-electrons which arises as a result of a combination between the crystal field, band filling and electron-electron correlation of $t_{2g}$-electrons, see e.g.~\cite{PhysRevLett.80.5758,PhysRevB.81.184432}. Our first-principles calculations of the AFM CoO for experimental lattice parameters predict the value of the gap of the order of 2.7\,eV, with the magnitude of local spin and orbital moments of 2.70\,$\mu_\mathrm{B}$ and 1.52\,$\mu_\mathrm{B}$ (see Materials and Methods S1 and Supplementary Text S6). The spin and orbital moments preserve their magnitude once the effect of spin-orbit coupling is switched off in electronic structure calculations.  While the spin degree of freedom is coupled to the orbital moments via  relatively weak spin-orbit interaction, the effect of significant orbital splitting among the states with opposite orbital polarization, which accounts for the order of the band gap, can be perceived in the orbital channel only as a result of an effective orbital exchange term in the Hamiltonian which signifies the repulsion energy between orbitally-polarized states. The presence of an effective strong orbital exchange in CoO allows one to reconsider the problem of interaction between the non-equilibrium OAM generated by Cu* and local orbital momenta of Co atoms in terms of an orbital torque, which acts solely within the orbital sector of quantum states. Given that the effect of orbital exchange in CoO is predominantly non-relativistic in nature, the corresponding dynamical orbital processes when out of equilibrium are expected to be much more manifest as compared to spin-orbit related dynamics hinging on a prior conversion of OAM into spin. This can explain a much stronger orbital current absorption by CoO in the Cu* system than that related to spin current-driven dynamics in Pt-based bilayers.

Another contribution to the giant OMR amplitude results from the conversion of OAM, generated by Cu*, corresponding to $L_{k}$, into a non-vanishing orbital position $\{L_i,L_j\}$ of the non-equilibrium states in CoO, for which $L_{i(j,k)}$ are cartesian components of the OAM operator. This should correspondingly give rise to a non-vanishing orbital quadrupole moment $\langle \{L_i,L_j\} \rangle$ inside CoO.  Recently, the currents of orbital quadrupole moments coupled to spin by spin-orbit interaction of Pt have been shown to mediate an injection of magnetic octupole into altermagnetic materials, thereby driving a novel type of torque on the Néel order parameter $\hat{\boldsymbol{n}}$  originated in the linear coupling between  $\hat{\boldsymbol{n}}$ and the magnetic octupole moment~\cite{han2024harnessingmagneticoctupolehall}. In CoO, it has been recently shown that the orbital quadrupole moment, arising as a combined result of an externally applied magnetic field and spin-orbit interaction, plays a key role in driving intricate spin-orbital dynamics via the term in energy which couples components of $\hat{\boldsymbol{n}}$ and orbital quadrupole moment $\langle \{L_x,L_y\} \rangle$~\cite{grzybowski2023antiferromagnetic}. We can show (see Materials and Methods S1 and Supplementary Text S6) that such a quadrupole moment can be created by injecting the 
component of the orbital angular momentum, which is  perpendicular to the equilibrium orientation of the spins and orbital moments. Due to the coupling with the spins, the induced quadrupole moment generates a torque on the N\'eel vector as well, thus inducing coupled spin-orbital  dynamics resulting in energy losses and an increase in the MR signal. 
Since the orbital dipole-to-quadrupole conversion does not in principle require spin-orbit interaction, we believe that the effective orbital quadrupole torques originating in the orbital injection from Cu* can be very efficient and ultimately result in strong absorption of the electrically-generated OAM.    

As a final remark, the demonstrated two-orders-of-magnitude enhancement in magnetoresistance in the CoO/Cu* system compared to conventional CoO/Pt can have major repercussions. Our findings suggest the potential for exploiting giant orbital currents for detecting and manipulating magnetic order through a novel orbital-orbital exchange mechanism. This mechanism arises from the combination of the giant orbital current generated in Cu* and the large unquenched static OAM in AFM CoO, enabling the development of all-orbital devices overcoming the limitations of conventional SOC mediated processes.


\clearpage 

%
\bibliography{science_template} 
\bibliographystyle{sciencemag}

%
%
%
%
%
%


\section*{Acknowledgments}
We acknowledge fruitful discussions with Dongwook Go, Daegeun Jo and Kyun-Jin Lee. We thank D. A. Grave, A. Kay and A. Rothschild for providing us with the $\alpha$-Fe$_2$O$_3$ thin film. The authors thank the DFG (Spin+X (A01, A11, B02) TRR 173-268565370 and Project No. 358671374), the Horizon 2020 Framework Programme of the European Commission under FETOpen Grant Agreement No. 863155 (s-Nebula); the European Research Council Grant Agreement No. 856538 (3D MAGiC); and the Research Council of Norway through its Centers of Excellence funding scheme, Project No. 262633 ”QuSpin”. We  also gratefully acknowledge the J\"ulich Supercomputing Centre  for providing computational resources under projects jiff40 and jiff38. The study also has been supported by the European Horizon Europe Framework Programme under an EC Grant Agreement N°101129641 ”OBELIX”. This research is part of the TOPOCOM project, which is funded by the European Union’s Horizon Europe Programme under the Marie Skłodowska-Curie Actions (MSCA), Grant Agreement No. 101119608. The work at the University of Tokyo was supported by CREST (Nos. JPMJCR20C1 and JPMJCR20T2) from JST, Japan, Grant-in-Aid for Scientific Research (S) (No. JP19H05600), Grant-in-Aid for Transformative Research Areas (No. JP22H05114) from JSPS KAKENHI, Japan, Grant-in-Aid for Scientific Research (B) (No. JP24K01326), Grant-in-Aid for Research Activity Start-up (No. 23K19024) from JSPS KAKENHI, Japan, Institute for AI and Beyond of the University of Tokyo, and Sumitomo Chemical.

\paragraph*{Author contributions:}
The CoO samples were grown by CS and LM with the assistance of TKi and HA supervised by ES. CS fabricated the devices and performed the electrical measurements. SK and EG assisted with the experiments.  CS and SK analyzed the data with inputs from TKu, YM and MK. ML, JS, OG and YM provided the theoretical support. DT helped in the sputtering of metal layers. CS, SK, YM and MK wrote the manuscript. MK proposed and supervised the whole project. All authors commented on the manuscript.

\paragraph*{Competing interests:}
The authors declare no competing interests.

\paragraph*{Data and materials availability:}
The data is available from the corresponding author on reasonable request.




\subsection*{Supplementary materials}
Materials and Methods\\
Supplementary Text\\
Figs. S1 to S6\\
References \textit{(57-\arabic{enumiv})}\\ 


\newpage


\renewcommand{\thefigure}{S\arabic{figure}}
\renewcommand{\thetable}{S\arabic{table}}
\renewcommand{\theequation}{S\arabic{equation}}
\renewcommand{\thepage}{S\arabic{page}}
\setcounter{figure}{0}
\setcounter{table}{0}
\setcounter{equation}{0}
\setcounter{page}{1} 


\begin{center}
\section*{Supplementary Materials for\\ \scititle}
    Christin~Schmitt,
	Sachin~Krishnia$^\ast$,
    Edgar~Gal\'{i}ndez-Ruales,
    Luca~Micus,
    Takashi~Kikkawa, 
    Hiroki~Arisawa, 
    Marjana~Le\v{z}ai\'c, 
    Duc~Tran, 
    Timo~Kuschel,
    Jairo~Sinova,
    Eiji~Saitoh,
    Gerhard~Jakob, 
    Olena~Gomonay, 
    Yuriy~Mokrousov,
    Mathias~Kl\"aui$^\ast$ \\
\small$^\ast$Corresponding authors. Email: skrishni@uni-mainz.de, klaeui@uni-mainz.de\\

\end{center}

\subsubsection*{This PDF file includes:}
Materials and Methods\\
Supplementary Text\\
Figures S1 to S6\\

\newpage


\section{Materials and Methods}

\subsection*{Sample growth and preparation}
This section describes the growth and structural characterization of our CoO thin films. The CoO($\SI{5}{nm}$) films were grown on MgO(001) substrates using reactive magnetron sputtering in an \textit{ULVAC QAM 4} fully automated sputtering system. The sputtering chamber was maintained at a base pressure better than $1.2 \times \SI{E-5}{Pa}$. Prior to deposition, the MgO substrates were pre-heated to a temperature of $\SI{770}{\degree C}$ for two hours. After cooling, CoO was deposited from a Co target in a mixed atmosphere of Ar ($\SI{15}{sccm}$) and $\mathrm{O_2}$ (\SI{2}{sccm}) at a temperature of $\SI{573}{\degree C}$ using an RF power of $\SI{150}{W}$. Following deposition and cooling down, non-magnetic metal layers such as Cu and Pt were deposited in-situ without breaking the vacuum at room temperature, utilizing a separate chamber and DC magnetron sputtering. The lattice constant and growth quality of CoO thin films are determined by using a Bruker D8 Discover high-resolution X-ray diffractometer. 

\subsection*{Device fabrication and magneto-transport measurements}

The as-grown samples were cleaned with acetone and isopropanol prior to patterning. The structures were patterned by photolithography using a DMO MicroWriter ML3 using negative photoresist and subsequent etching using Ar+ ion milling technique. The Hall bar geometry consists of a $\SI{6}{\micro m}$ wide bar and two $\SI{2}{\micro m}$ wide wires perpendicular to it, separated from each other by $\SI{6}{\micro m}$ patterned along the sample hard axes magnetic anisotropy. The Pt layer of the CoO($\SI{5}{nm}$)/Cu($\SI{6}{nm}$) sample was removed by argon etching for $\SI{8}{s}$. For the electrical measurements, the sample was mounted onto a piezo-rotating element within a variable temperature insert (VTI) installed in a superconducting coil capable of generating magnetic fields up to $\SI{15}{T}$. 
For field-sweep experiments, the magnetic field was swept in-plane along the two perpendicular magnetic easy axes. For angular-dependent measurements, a constant magnetic field was rotated in-plane.
A charge current of $\SI{105}{\micro A}$ was applied to the  $\SI{6}{\micro m}$ wide bar using a Keithley 6221 current source, alternating between positive and negative polarity. The resulting transverse and longitudinal voltage signals were detected in two transverse bars using a Keithley 2182A nanovoltmeter. The voltage value is an average of 100 measurements, with each measurement being the average of positive and negative current polarity. The error bar is the standard deviation of these measurements. The applied current followed a square-wave signal as a function of time. A delay was introduced between the current reversal and the start of data acquisition to avoid capturing rise-time effects from the current source or capacitor-like effects in the Hall structures. By calculating the difference between the two signals arising from positive and negative current polarities, thermal contributions, such as the spin Seebeck effect, are averaged out. The resulting voltage variation was recorded as a function of the externally applied magnetic field and the angle between the charge current and the magnetic field.

\subsection*{First-principles calculations} We performed our density functional theory (DFT) calculations of CoO using the full-potential linearized augmented plane wave method (FLAPW), as implemented in the Jülich DFT code FLEUR \cite{fleur}, with the Perdew-Burke-Ernzerhof \cite{Perdew1996} parametrization of the exchange-correlation potential.  For the self-consistent calculations, a 16$\times$16$\times$16 $k$-point mesh was used to sample the whole Brillouin zone.  To account for the effect of electron-electron correlation among 3$d$-electrons of Co, the GGA+$U$ method in Liechtenstein approach was applied within a self-consistent DFT cycle. The on-site Coulomb and Hund exchange parameters were set to $U$ = 6.0 eV and $J$ = 0.92 eV.
We set the muffin-tin radii to 2.48 a.u. for Co and 1.4 a.u. for oxygen atoms. The cut-off for the plane-wave basis functions was chosen as $K_{\text{max}}$ = 4.3 and $G_{\text{max}}$ = 12.9\,a.u.$^{-1}$ for the charge density and potential. The upper limit of angular momentum inside of the muffin-tin spheres was set to $l_{\text{max}}$ = 10 and 6 for Co and O respectively.

\begin{figure}[h]
\centering
\includegraphics[width=0.9\textwidth]{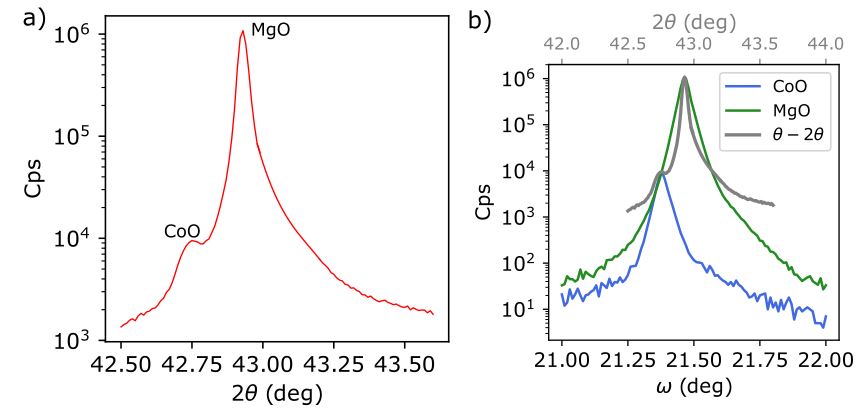}
\caption{\textbf{Structural properties of CoO thin films grown on MgO(001) substrate.} (a) $\theta-2\theta$ XRD scan showing the (001) alignment of the CoO($\SI{5}{nm}$)/Cu($\SI{6}{nm}$) film on MgO(001) substrate. (b) XRD rocking-curves of CoO(002) and MgO(002) planes on top of the $\theta-2\theta$ scan (grey) from panel (a).}\label{XRD}
\end{figure}

\section{Characterization of structural and crystallographic properties of CoO thin films}

To determine the crystallographic orientation, lattice constant  and  the growth quality of our CoO thin films, we performed X-ray diffraction (XRD) measurements using a Bruker D8 Discover high-resolution diffractometer with Cu $K_\alpha$ radiation ($\lambda = \SI{0.15406}{nm}$). Fig. \ref{XRD}(a) shows the coupled $\theta-2\theta$ scan for MgO(001)//CoO($\SI{5}{nm}$)/Cu($\SI{6}{nm}$). The observed peak position for CoO is consistent with the CoO(002) direction, in alignment with the orientation of the MgO(001) substrate. The position of CoO(002) is slightly shifted relative to the expected bulk value, suggesting that the CoO thin film is strained. This strain arises from a lattice mismatch of 1.1\% between the bulk lattice constants of MgO substrate ($a = \SI{0.4212}{nm}$) and CoO ($a = \SI{0.4260}{nm}$). Fig. \ref{XRD}(b) presents the rocking curves for the CoO(002) thin film and MgO(002) substrate peaks. Both curves exhibit similar full-width-at-half-maximum (FWHM) values, indicating the high crystalline quality of the CoO film, with minimal dislocations, mosaicity, and defects. Moreover, the CoO peak is centered at the expected position of the film's Bragg reflection, highlighting the excellent lattice alignment between the film and substrate.

\section{Measurements on $\mathrm{\alpha-Fe_2O_3/Cu^*}$}

\begin{figure}[h]
\centering
\includegraphics[width=0.9\textwidth]{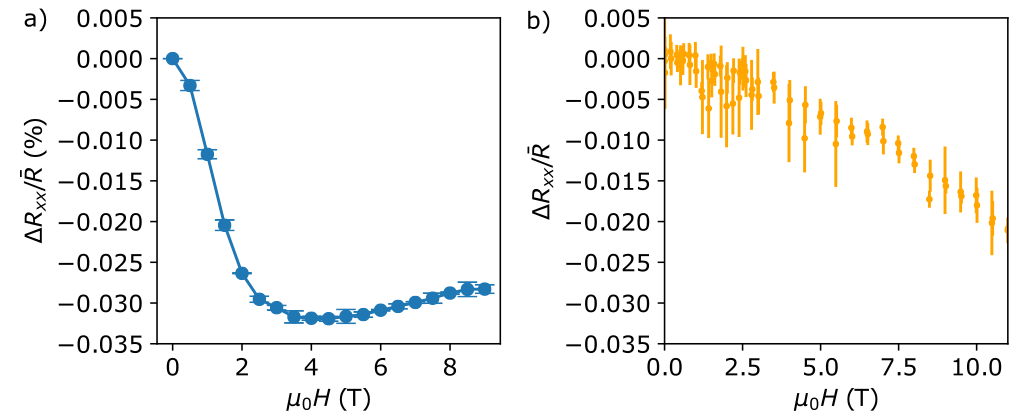}
\caption{\textbf{Interaction of spin and orbital currents with spin-dominated AFM $\mathbf{\mathrm{\alpha-Fe_2O_3}}$.} Longitudinal resistance of hematite/ (a) Pt and (b) Cu* bilayers at $\SI{175}{K}$ in the easy-axis phase for a magnetic field applied along the current direction.} \label{fig:hematite}
\end{figure}

To confirm that the observed effects in CoO/Cu* are originating from orbital physics, we performed MR measurements on $\mathrm{\alpha-Fe_2O_3(hematite)/Cu^*}$. Hematite has a nearly quenched OAM and is thus expected to show negligible response to the orbital currents \cite{rollmann2004first}. For this, we used a $\SI{100}{nm}$ thin film in c-cut orientation, interfaced with $\SI{5}{nm}$ thick Cu*.
On this particular $\mathrm{\alpha-Fe_2O_3}$ sample, we performed SMR measurements using a spin current generated in Pt, as shown in Fig.~\ref{fig:hematite}(a). We found that, due to the structure of this sample, SMR of the order of $\sim3 \times 10^{-4}$ arises at low magnetic field when performing field-sweep measurements at $\SI{175}{K}$ in the easy-axis phase (see also Ref.~\cite{ross2020structural}). 

We patterned Hall bars equivalent to those used for the measurements on CoO and injected electrical currents up to $\SI{1}{mA}$, to ensure a good signal-to-noise ratio. We applied the magnetic field parallel to the direction of the electric current. We observed no significant change in the resistance at low magnetic fields (up to $\SI{2}{T}$). Only at high magnetic fields, a resistance change in the order of ${\sim2 \times 10^{-4}}$ is observed. However, as this magnetic field strength is far above the spin-flop field of $\mathrm{\alpha-Fe_2O_3/Cu^*}$, the origin of this signal cannot be magnetic from the hematite. Thus, we confirm that the observed giant OMR in CoO/Cu* originates from the interaction between the orbital current and the unquenched OAM of CoO.

\section{Cu thickness determination}
\begin{figure}[h]
\centering
\includegraphics[width=0.9\textwidth]{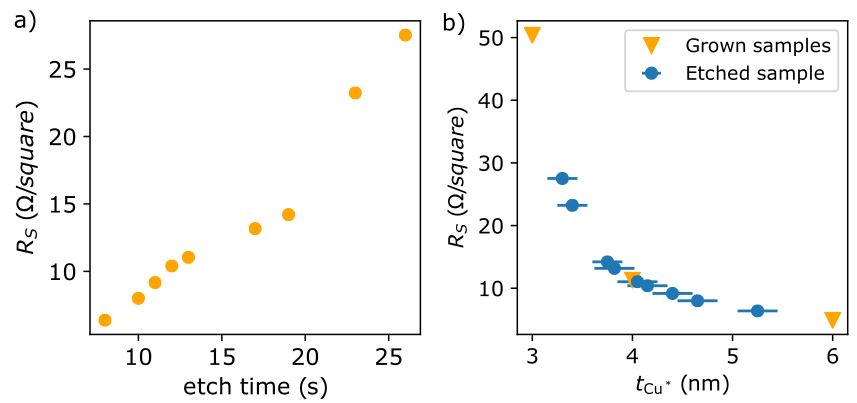}
\caption{\textbf{Thickness calibration of etched Cu* using resistivity measurements.}(a) Sheet resistance $R_S$ of the etched Hall bar device as a function of etch time. (b) Comparison of sheet resistance of as-grown films with varying Cu* thickness and the sheet resistance of a Cu* layer after several Ar-ion etching steps.} \label{sheetResistance}
\end{figure}

To ensure maximum comparability of the Cu* thickness dependence on OMR, all Cu* thickness dependence measurements were performed on the same Hall bar device. This approach minimizes the impact of any potential variation arising from the CoO/Cu* interface due to growth conditions on the measured OMR. The Cu* thickness was progressively reduced by using repeated Ar-ion etching of the Hall bar device used for the electrical measurements to study the OMR. After each etching step, the precise  Cu* thickness was determined by measuring the sheet resistance of the Cu* layer and comparing it with the sheet resistances of as-grown films of known Cu* thicknesses. For the as-grown samples with varying Cu* thicknesses, we used the Van der Pauw method \cite{pauw1958method} to determine the sheet resistance ($R_s$). For the etched Hall bar structure, we measured the longitudinal resistance at room temperature and accounted for the device geometry
($w = \SI{8}{nm},\ l = \SI{6}{nm}$) by dividing it by a factor of 1.33. Fig. \ref{sheetResistance} (a) shows the sheet resistance as a function of etch time. In Fig. \ref{sheetResistance} (b), we show how the sheet resistances of the etched films were matched to those of the as-grown samples. It is worth noting that a Cu* layer with  a thickness of $\SI{2}{nm}$ was already fully oxidized, rendering the sheet resistance measurement impossible.

\section{Angular dependent measurements on CoO/Pt and CoO/Cu*}

\begin{figure}[h]
\centering
\includegraphics[width=0.85\textwidth]{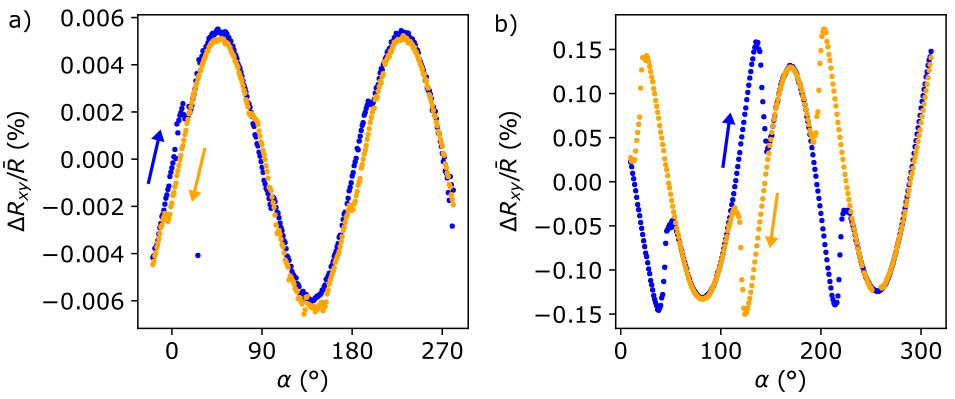}
\caption{\textbf{Angular dependence of transverse spin and orbital magnetoresistance in CoO.} Angular dependence of transverse magnetoresistance in (a) CoO/Pt and (b) CoO/Cu* at $\SI{200}{K}$ for an applied magnetic field of $\SI{11}{T}$. The blue and orange data points represent two different rotation directions, as indicated by the coloured arrows.} \label{fig:anglesweep}
\end{figure}

We performed angular-dependent magnetoresistance measurements on CoO/Pt and CoO/Cu* at various external magnetic fields, ranging from below  to above the spin-flop fields of CoO. A static magnetic field of up to $\SI{11}{T}$ was applied at a temperature of $\SI{200}{K}$, and the sample was rotated in the \textit{x-y} plane to vary the in-plane angle between the applied electric current and the magnetic field. The angle was varied over a range of $\SI{300}{\degree}$ in steps of $\SI{1}{\degree}$. The MR of CoO/Pt and CoO/Cu* is shown in Fig. \ref{fig:anglesweep} (a) and (b), respectively.

\begin{figure}[h]
\centering
\includegraphics[width=0.9\textwidth]{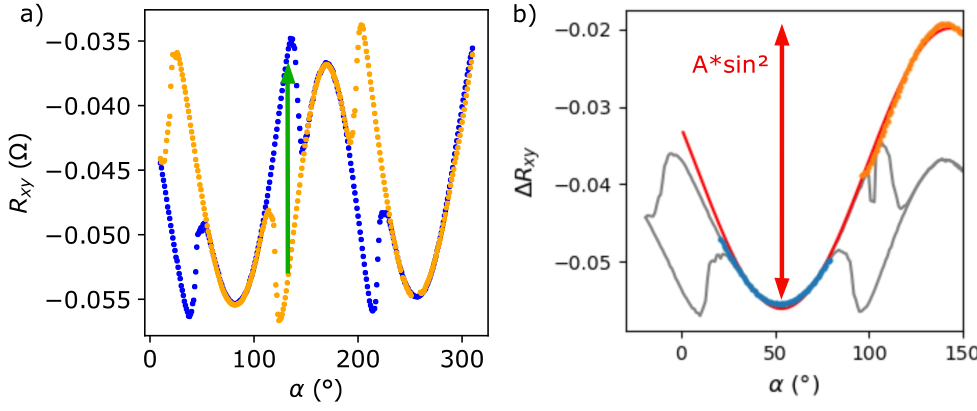}
\caption{\textbf{Disentangling magnetic and spurious contribution to OMR in CoO/Cu*}. Exemplary data of CoO/Cu* to show how the two contributions to the signal can be disentangled. (a) The amplitude of the hysteretic signal can be determined by shifting one angle sweep direction (orange) of the hysteresis to the second one (blue), as shown by the green arrow. (b) By fitting a $sin^2(x)$-function (red) to the minimum and maximum for positive and negative sweep direction (blue and orange) the amplitude of the sinusoidal signal is determined. } \label{fig:disentangle}
\end{figure}

\begin{figure}[h]
\centering
\includegraphics[width=0.35\textwidth]{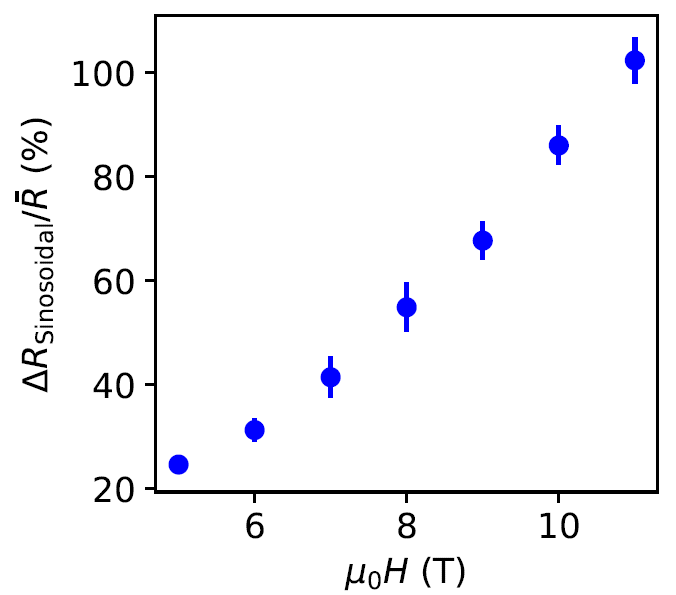}
\caption{\textbf{Field dependence of spurious contribution to OMR in CoO/Cu*.} Amplitudes of the sinusoidal signal as a function of applied magnetic field.} \label{fig:amplitudes}
\end{figure}

Our measurements reveal that, for externally applied magnetic fields below the spin-flop field ($\SI{7}{T}$), the angular scans in the $x$-$y$ plane exhibit a sinusoidal variation in transverse resistance. However, when the in-plane magnetic field exceeds the spin-flop field, additional hysteresis loops emerge, centered around the hard-axis directions ([100], [010], and $[\bar{1}00]$) of the sample. These hysteresis loops can be attributed to field-induced anisotropy arising from the unquenched OAM in CoO \cite{grzybowski2023antiferromagnetic}. The different behaviour of the two signals raises questions about their origin and their respective contributions to the overall MR. To address this, we analyzed the magnetic field dependence of the transverse resistance by separating the sinusoidal and hysteretic components. We first calculated the required shift to align one branch of the hysteresis curve with the other. The magnitude of this shift corresponds to the amplitude of the hysteretic signal. Subsequently, we fitted a $\mathrm{sin}^2(x)$ function to the signal to represent the data without the hysteretic contribution, thereby determining the amplitude of the sinusoidal component. The two steps are illustrated in Fig.~\ref{fig:disentangle}.
As shown in Fig.~\ref{fig3} of the manuscript, we find that the hysteretic signal emerges only above the spin-flop field and saturates at $\SI{9}{T}$. In contrast, the sinusoidal contribution increases linearly with the applied magnetic field as shown in Fig.~\ref{fig:amplitudes}. The magnetic field range for which a hysteretic signal is observed without reaching the saturated amplitude yet, corresponds well to the range of magnetic field strengths in Fig. 2(b) of the manuscript for which the spin flop transition is observed, thus indicating that the hysteresis in the MR signal corresponds to the switching of the Néel vector. The magnitude of the hysteretic component quantifies the MR amplitude. At $\SI{200}{K}$, the extracted MR values for the CoO/Cu* and CoO/Pt samples are 0.25$\%$ and 0.002$\%$, respectively. These values are consistent with those obtained using the zero-field approach (see Fig.~\ref{fig2}c). Consequently, the observed hysteresis in the transverse MR not only confirms the presence of significant unquenched OAM in our samples but also provides a reliable measure of the MR amplitude through the height of the hysteretic component. On the other hand, the sinusoidal signal likely arises as an artifact due to MR contributions from the Cu layer and is not of magnetic origin.

\section{Theoretical model}
The orbital state of CoO and coupling between the spins and orbital moments are described with the hamiltonian
\begin{equation}\label{eq-hamiltonian}
\hat{\mathcal{H}}=\sum_{j=1,2}\left[-K^L\left(\hat{\mathbf{L}}_j\mathbf{e}_j\right)^2-K^\mathrm{ea}_\mathrm{an}\left(\hat{L}_{jx}^2-\hat{L}_{jy}^2\right)\left(S_{jx}^2-S_{jy}^2\right)+K^\mathrm{ha}_\mathrm{an}\left\{\hat{L}_{jx},\hat{L}_{jy}\right\}S_{jx}S_{jy}+\lambda\mathbf{S}_j\hat{\mathbf{L}}_j\right].
\end{equation}
Here the $\mathbf{S}_j$ ($j=1,2$) are the magnetic spins at two magnetic sublattices, treated as the classical vectors. Operators $\hat{\mathbf{L}}_j$  describe the effective orbital angular momenta ($L=1$) at the Co sites. The first term in Eq.\eqref{eq-hamiltonian} describes anisotropy of the angular momenta. The local quantization axes $\mathbf{e}_1\downarrow\downarrow\mathbf{e}_2\uparrow\uparrow\hat{x}$ are fixed along the equilibrium orientation of the N\'eel vector and opposite to the sublattice magnetizations. Next two terms describe contribution into magnetic anisotropy with values  parametrized by the constants $K^\mathrm{ea}_\mathrm{an},K^\mathrm{ha}_\mathrm{an}>0$. The last term in Eq.\eqref{eq-hamiltonian} describes spin-orbit coupling with $\lambda>0$ being the coupling constant.  Orthogonal axes $x$ and $y$ are parallel to the in-plane easy magnetic axes [110] and [1$\bar{1}$0].

Spin dynamics of CoO is described with the standard LLG equations for the N\'eel vector $\mathbf{n}=\mathbf{S}_1-\mathbf{S}_2$:
\begin{equation}
    \mathbf{n}\times\ddot{\mathbf{n}}=-\gamma^2H_\mathrm{ex}\mathbf{n}\times\frac{\partial \langle\hat{\mathcal{H}}\rangle}{\partial \mathbf{n}},
\end{equation}
where $\langle \ldots\rangle$ denotes the quantum-mechanical average over the orbital degrees of freedom, $\gamma$ is gyromagentic ratio, $H_\mathrm{ex}$ is intersublattice exchange field that keeps sublattice spins antiparallel.

\end{document}